\documentclass{llncs}

\usepackage{macros}
\usepackage{lstcoq}
\usepackage{lstocaml}
\usepackage{mathpartir}
\usepackage{amsfonts}
\usepackage{amsmath}
\usepackage{stmaryrd}
\usepackage{graphics}
\usepackage{array}

\author{Thomas Braibant$^1$ \and Adam Chlipala${}^{2}$}
\institute{${}^1$ Inria \qquad ${}^2$ MIT} %
\title{Formal Verification of Hardware Synthesis}
\newcommand{\project}{Fe-Si}

\newcommand{\denote}[1]{\llbracket #1 \rrbracket}
\newcommand{\denotety}[1]{\denote{\mathtt{#1}}_{\mathtt{ty}}}
\newcommand{\denotemem}[1]{\denote{\mathtt{#1}}_{\mathtt{mem}}}

\begin{document}
\maketitle

\begin{abstract}
  We report on the implementation of a certified compiler for a
  high-level hardware description language (HDL) called \emph{Fe-Si}
  (FEatherweight SynthesIs).
  Fe-Si is a simplified version of Bluespec, an HDL based on a notion
  of \emph{guarded atomic actions}. Fe-Si is defined as a
  dependently typed deep embedding in Coq. The target language of the
  compiler corresponds to a synthesisable subset of Verilog or VHDL.
  A key aspect of our approach is that input programs to the compiler
  can be defined and proved correct inside Coq. Then, we use
  extraction and a Verilog back-end (written in OCaml) to get a
  certified version of a hardware design.
\end{abstract}

\section*{Introduction}
Verification of hardware designs has been thoroughly investigated, and
yet, obtaining provably correct hardware of significant complexity is
usually considered challenging and time-consuming. 
On the one hand, a common practice in hardware verification is to take
a given design written in an hardware description language like
Verilog or VHDL and argue about this design in a formal way using a
model checker or an SMT solver.
On the other hand, a completely different approach is to design
hardware via a shallow embedding of circuits in a theorem
prover~\cite{hanna-veritas,UCAM-CL-TR-77,hunt89,vamp,certifying-circuits-in-type-theory}.
Yet, both kinds of approach suffer from the fact that most hardware
designs are expressed in low-level register transfer languages (RTL)
like Verilog or VHDL, and that the level of abstraction they provide
may be too low to do short and meaningful proof of high-level
properties.

\medskip

To raise this level of abstraction, industry moved to \emph{high-level
  hardware synthesis} using higher-level languages, e.g.,
System-C~\cite{systemc}, Esterel~\cite{DBLP:conf/birthday/Berry00} or
Bluespec~\cite{bluespec}, in which a source program is
compiled to an RTL description.
High-level synthesis has two benefits. 
First, it reduces the effort necessary to produce a hardware design.
Second, writing or reasoning about a high-level program is simpler
than reasoning about the (much more complicated) RTL description
generated by a compiler.
However, the downside of high-level synthesis is that there is no
formal guarantee that the generated circuit description behaves
exactly as prescribed by the semantics of the source
program, making verification on the source program useless in the
presence of compiler-introduced bugs.

\medskip In this paper, we investigate the formal verification of a
lightly optimizing compiler from a Bluespec-inspired language called
\project{} to RTL, applying (to a lesser extent) the ideas behind the
CompCert project~\cite{Leroy-backend} to hardware synthesis.

\medskip

\project{} can be seen as a stripped-down and simplified version
of Bluespec: in both languages, hardware designs are described in
terms of \emph{guarded atomic actions} on state elements. 
The one oddity here is that this language and its semantics have a
flavor of \emph{transactional memory}, where updates to state
elements are not visible before the end of the transaction (a
time-step).
Our target language can be sensibly interpreted as \emph{clocked
  sequential machines}: we generate an RTL description syntactically
described as combinational definitions and next-state assignments.
In our development, we define a (dependently typed) deep embedding of
the \project{} programming language in Coq using \emph{parametric
  higher-order abstract syntax (PHOAS)}~\cite{phoas-chlipala}, and
give it a semantics using an interpreter: the semantics of a program
is a Coq function that takes as inputs the current state of the
state elements and produces a list of updates to be committed to
state elements.

\project{} hits a sweet spot between deep and shallow embeddings: using
PHOAS to embed a domain-specific language makes it possible to use Coq
as a meta programming tool to describe circuits, without the pitfalls
usually associated with a deep embedding (e.g., the handling of
binders).
This provides an economical path toward succinct and provably correct
description of, e.g., recursive circuits.

\section{Overview of Fe-Si}
Fe-Si is a purely functional language built around a \emph{monad} that
makes it possible to define circuits. We start with a customary
example: a half adder.
\begin{mcoq}
Definition hadd (a b: Var B) : action [] (B $\otimes$ B) :=
$\quad$do carry <- ret (andb a b); 
$\quad$do sum    <- ret (xorb a b);
$\quad$ret (carry, sum).  
\end{mcoq}
This circuit has two Boolean inputs (\coqe{Var B}) and return a tuple
of Boolean values (\coqe{B $\otimes$ B}).
Here, we use Coq notations to implement some syntactic sugar: we
borrow the \texttt{do}-notation to denote the monadic bind and use
\coqe{ret} as a short-hand for return. 
(Our choice of concrete notations is dictated by some limitations in
Coq's notation mechanism. For instance our explicit use of return may
seem odd: it is due to the fact that Fe-Si has two classes of
syntactic values, expressions and actions, and that return takes as
argument an expression.)

Up to this point, Fe-Si can be seen as an extension of the
Lava~\cite{Bjesse98lava:hardware} language, implemented in Coq rather
than Haskell. Yet, using Coq as a metalanguage offers the possibility
to use dependent types in our circuit descriptions. For instance, one
can define an adder circuit of the following type:
\begin{mcoq}
Definition adder n (a b: Var (Int n)): action [] (Int n) := ...
\end{mcoq}
In this definition, \coqe{n} of type \coqe{nat} is a formal parameter
that denotes the size of the operands and the size of the result as
well. 

\subsubsection{Stateful programs.}
Fe-Si also features a small set of primitives for interacting with
\emph{memory elements} that hold mutable state. In the following
snippet, we build a counter that increments its value when its input
is true.
\begin{mcoq}
Definition $\Phi$ := [Reg (Int n)].
Definition count n (tick: Var B) : action $\Phi$ (Int n) :=
$\quad$do x <- !member_0;
$\quad$do _ <- if tick then {member_0 ::= x + 1} else {ret ()}; 
$\quad$ret x. 
\end{mcoq}
Here, $\Phi$ is an environment that defines the set of memory elements
(in a broad sense) of the circuit. In the first line, we read the
content of the register at position \coqe{member_0} in $\Phi$, and
bind this value to \coqe{x}. Then, we test the value of the input
\coqe{tick}, and when it is true, we increment the value of the
register. In any case, the output is the old value of the counter.

The above ``if-then-else'' construct is defined using two primitives
for guarded atomic actions that are reminiscent of transactional
memory monads: \coqe{assert} and \coqe{orElse}. The former aborts the
current action if its argument is false.
The latter takes two arguments $a$ and $b$, and first executes $a$; if
it aborts, then the effects of $a$ are discarded and $b$ is run. If
$b$ aborts too, the whole action \coqe{$a$ orElse $b$} aborts.

\subsubsection{Synchronous semantics.} Recall that Fe-Si programs are
intended to describe hardware circuits. Hence, we must stress that
they are interpreted in a synchronous setting.
From a logical point of view the execution of a program (an atomic
action) is clocked, and at each tick of its clock, the computation of
its effects (i.e., updates to memory elements) is instantaneous: these
effects are applied all at once between ticks.
In particular this means that it is not possible to observe, e.g.,
partial updates to the memory elements, nor transient values in
memory.
(In Bluespec terminology, this is ``reads-before-writes''.)

\subsubsection{From programs to circuits.} At this point, the reader
may wonder how it is possible to generate circuits in a palatable
format out of Fe-Si programs. Indeed, using Coq as a meta-language to
embed Fe-Si yields two kind of issues. First, Coq lacks any kind of
I/O; and second, a Fe-Si program may have been built using arbitrary
Coq code, including, e.g., higher-order functions or fixpoints.

Note that every Coq function terminates: therefore, a closed Fe-Si
program of type \coqe{action} evaluates to a term that is
syntactically built using the inductive constructors of the type
\coqe{action} (i.e., all intermediate definitions in Coq have been expanded).
Then we use Coq's extraction, which generates OCaml code from Coq
programs.  
Starting from a closed Fe-Si program \coqe{foo}, we put the following
definition in a Coq file:
\begin{mcoq}
Definition bar := fesic foo.  
\end{mcoq}
The extracted OCaml term that corresponds to \coqe{bar} evaluates (in
OCaml) to a closed RTL circuit. Then, we can use an (unverified)
back-end that pretty-prints this RTL code as regular Verilog code.
(We reckon that this is some devious use of the extraction mechanism,
which palliates the fact that there is currently no I/O mechanism in
Coq.)

\section{From Fe-Si to RTL}

In this section, we shall present our source (Fe-Si) and target (RTL)
languages, along with their semantics. For the sake of space, we leave
the full description of this compilation process out of the scope of
this paper.

\subsection{The memory model}
Fe-Si programs are meant to describe sequential circuits, whose
``memory footprints'' must be known statically. We take a declarative
approach: each state-holding element that is used in a program must be
declared. 
We currently have three types of memory elements: inputs, registers,
and register files. A register holds one value of a given type, while a
register file of size $n$ stores $2^n$ values of a given type. 
An input is a memory element that can only be read by the circuit,
and whose value is driven by the external world.
We show the inductive definitions of types and memory elements in
Fig.~\ref{fig:type}. 
We have four constructors for the type \coqe{ty} of types: \coqe{Unit}
(the unit type), \coqe{B} (Booleans), \coqe{Int} (integers of a given
size), and \coqe{Tuple} (tuples of types). The inductive definition of
memory elements (\coqe{mem}) should be self-explaining. 

We endow these inductive definitions with a denotational semantics: we
implement Coq functions that map such reified types to the obvious Coq
types they denote.

\begin{figure}
  \centering
\begin{threelistings}
\begin{coq}
Inductive ty : Type :=
| Unit : ty 
| B : ty 
| Int : nat -> ty
| Tuple : list ty -> ty.     
\end{coq}&
\begin{coq}
Inductive mem : Type :=
| Input: ty ->  mem
| Reg : ty -> mem
| Regfile : nat -> ty -> mem. 
$ $
\end{coq}
&
\begin{coq}
Fixpoint $\denotety{.}$ : ty -> Type := ...
Fixpoint $\denotemem{.}$ : mem -> Type := ...
Fixpoint $\denote{.}$ : list mem -> Type := ...

$ $
\end{coq}
\end{threelistings}
\caption{Types and memory elements}
  \label{fig:type}
\end{figure}

\subsection{Fe-Si}
The definition of Fe-Si programs (\coqe{action} in the following)
takes the PHOAS approach~\cite{phoas-chlipala}. 
That is, we define an inductive type family parameterized by an
arbitrary type \coqe{V} of variables, where binders bind variables
instead of arbitrary terms (as would be the case using
HOAS~\cite{DBLP:conf/pldi/PfenningE88}), and those variables are used
explicitly via a dedicated term constructor.
The definition of Fe-Si syntax is split in two syntactic classes:
expressions and actions. 
Expressions are side-effect free, and are built from variables,
constants, and operations.
Actions are made of control-flow structures (assertions and
alternatives), binders, and memory operations. 

In this work, we follow an intrinsic
approach~\cite{DBLP:journals/jar/BentonHKM12}: we mix the definition
of the abstract syntax and the typing rules from the start. That is,
the type system of the meta-language (Coq) enforces that all Fe-Si
programs are well-typed by construction.
Besides the obvious type-oblivious definitions (e.g., it is not
possible to add a Boolean and an integer), this means that the
definition of operations on state-holding elements requires some care.
Here, we use dependently typed de Bruijn indices. 
\begin{mcoq}
Inductive member : list mem -> mem ->  Type :=
| member_0 : forall E t, member (t::E) t
| member_S : forall E t x, member E t -> member (x::E) t.
\end{mcoq}
Using the above definition, a term of type \coqe{member $\Phi$ M} denotes
the fact that the memory element \coqe{M} appears at a given position
in the environment of memory elements $\Phi$. 
We are now ready to present the (elided) Coq definitions of the
inductives for expressions and actions in Fig.~\ref{fig:fesi}.
(For the sake of brevity, we omit the constructors for accesses to
register files, in the syntax and, later, in the semantics. We refer
the reader to the supplementary material~\cite{fesi} for more details.)
Our final definition \coqe{Action} of actions is a polymorphic
function from a choice of variables to an action (we refer the reader
to \cite{phoas-chlipala} for a more in-depth explanation of this
encoding strategy).

\begin{figure}[t]
  \centering
\begin{coq}
Section t. 
  Variable V: ty -> Type. Variable $\Phi$: list mem. 
  Inductive expr: ty -> Type :=
  | Evar : forall t (v : V t), expr t
  (* operations on Booleans *)
  | Eandb : expr B -> expr B -> expr B | ... 
  (* operations on words *)
  | Eadd : forall n, expr (Int n) -> expr (Int n) -> expr (Int n) | ... 
  (* operations on tuples *)
  | Efst : forall l t, expr (Tuple (t::l)) -> expr t | ...

  Inductive action: ty -> Type:=
  | Return: forall t, expr t -> action t
  | Bind: forall t u,  action  t -> (V t -> action u) -> action u
  (* control-flow *)
  | OrElse: forall t, action t -> action t -> action t
  | Assert: expr B -> action Unit    
  (* memory operations on registers *)
  | RegRead : forall t, member $\Phi$ (Reg t) -> action t
  | RegWrite: forall t, member $\Phi$ (Reg t) -> expr t -> action Unit
  (* memory operations on register files, and inputs *)
  | ... 
End t. 
Definition Action $\Phi$ t := forall V, action V $\Phi$ t.  
\end{coq}
  \caption{The syntax of expressions and actions}
  \label{fig:fesi}
\end{figure}

\subsubsection{Semantics.}
We endow Fe-Si programs with a simple synchronous semantics:  starting
from an initial state, the execution of a Fe-Si program corresponds
to a sequence of atomic updates to the memory elements. 
Each step goes as follows: reading the state, computing an update to
the state, committing this update.

\begin{figure*}
  \centering
  \begin{mathpar}
    \inferrule{\Gamma \vdash e \leadsto v} {\Gamma, \Delta \vdash
      \mathtt{Return}~e \to \mathtt{Some} (v,\Delta)}
    \\
    \inferrule{ \Gamma, \Delta_1 \vdash a \to \mathtt{None} } {\Gamma,
      \Delta_1 \vdash \mathtt{Bind}~a~f \to \mathtt{None}} \and
    \inferrule{ \Gamma, \Delta_1 \vdash a \to \mathtt{Some}~(v,
      \Delta_2) \and \Gamma, \Delta_2 \vdash f~v \to r} {\Gamma,
      \Delta_1 \vdash \mathtt{Bind}~a~f \to r}
    \\
    \inferrule{\Gamma \vdash e \leadsto \mathtt{true}} {\Gamma, \Delta
      \vdash \mathtt{Assert}~e \to \mathtt{Some} (\mathtt{()},\Delta)}
    \and \inferrule{\Gamma \vdash e \leadsto \mathtt{false}} {\Gamma,
      \Delta \vdash \mathtt{Assert}~e \to \mathtt{None}}
    \\
    \inferrule{ \Gamma, \Delta \vdash a \to \mathtt{Some}~(v,\Delta')}
    {\Gamma, \Delta \vdash a~\mathtt{orElse}~b \to
      \mathtt{Some}~(v,\Delta')} \and
    \inferrule{\Gamma, \Delta \vdash a \to \mathtt{None} \and \Gamma,
      \Delta \vdash b \to r}
    {\Gamma, \Delta \vdash a~\mathtt{orElse}~b \to r}
    \\
    \inferrule{\Gamma(r) = v} {\Gamma, \Delta \vdash
      \mathtt{RegRead}~r \to \mathtt{Some} (r,\Delta)} \and
    \inferrule{\Gamma \vdash e \leadsto v} {\Gamma, \Delta \vdash
      \mathtt{RegWrite}~r~e \to \mathtt{Some}
      (\mathtt{()},\Delta\oplus(r,v))}
        
  \end{mathpar}
\caption{Dynamic semantics of Fe-Si programs}\label{fig:fesi-sem}
\end{figure*}

The reduction rules of Fe-Si programs are defined in
Fig.~\ref{fig:fesi-sem}. The judgement $\Gamma, \Delta \vdash a \to r$
reads ``in the state $\Gamma$ and with the partial update $\Delta$,
evaluating $a$ produces the result $r$'', where $r$ is either
\coqe{None} (meaning that the action aborted), or %
\coqe{Some (v, $\Delta'$)} (meaning that the action returned the value
\coqe{v} and the partial update $\Delta'$). 
Note that the PHOAS approach makes it possible to manipulate closed
terms: we do not have rules for $\beta$-reduction, because it is
implemented by the host language.
That is, $\Gamma$ only stores the mutable state, and not the variable
values.
There are two peculiarities here: first, following the definition of
$\oplus$, if two values are written to a memory element, only the
first one (in program order) is committed; second, reading a register
yields the value that was held at the beginning of the time step. 

(The reduction rules of Fe-Si can be described in terms of layered
monads: we have a Reader monad of the old state and Writer monad of
the state update to implement the synchronous aspect of state update;
and we stack the Option monad on top of the state-change monads to
implement the transactional aspect of the semantics.)

Finally, we define a wrapper function that computes the next state of
the memory elements, using the aforementioned evaluation relation
(starting with an empty partial update). 
\begin{mcoq}
Definition Next {t} {$\Phi$} (st: $\denote{\Phi}$) (A : Action $\Phi$ t) : option ($\denotety{t}$ * $\denote{\Phi}$) := ...
\end{mcoq}

\subsection{RTL} 
Our target language sits at the register-transfer level. At this
level, a synchronous circuit can be faithfully described as a set of
state-holding elements, and a next-state function, implemented using
combinational logic~\cite{DBLP:journals/cj/Gordon02}.
Therefore, the definition of RTL programs (\coqe{block} in the
following) is quite simple: a program is simply a well-formed sequence
of bindings of expressions (combinational operations, or reads from
state-holding elements), with a list of effects (i.e, writes to
state-holding elements) at the end.

We show the definition of expressions and sequences of binders in
Fig.~\ref{fig:rtl}.
The definition of expressions is similar to the one we used for Fe-Si,
except that we have constructors for reads from memory elements, and
that we moved to ``three-adress code''.
(That is, operands are variables, rather than arbitrary expressions.)
A telescope (type \coqe{scope A}) is a well-formed sequence of binders
with an element of type \coqe{A} at the end (\coqe{A} is instantiated
later with a list of effects). Intuitively, this definition enforces
that the first binding of a telescope can only read from memory
elements; the second binding may use the first value, or read from
memory elements; and so on and so forth.

A \coqe{block} is a telescope, with three elements at the end: a
guard, a return value, and a (dependently typed) list of effects. 
The value of the guard (a Boolean) is equal to true when the return
value is valid and the state updates must be committed; and false
otherwise.
The return value denotes the outputs of the circuits. 
The data type \coqe{effects} encodes, for each memory element of the
list $\Phi$, either an effect (a write of the right type), or \coqe{None}
(meaning that this memory element is never written to). (For the sake
of brevity, we omit the particular definition of dependently typed
heterogeneous lists \coqe{DList.T} that we use here.)

\begin{figure}[t]
  \centering
\begin{coq}
Section t. 
  Variable V: ty -> Type. Variable $\Phi$: list mem. 
  Inductive expr: ty -> Type :=
  | Evar : forall t (v : V t), expr t
  (* read from memory elements *)
  | Einput : forall t, member $\Phi$ (Input t) -> expr t
  | Eread_r : forall t, member $\Phi$ (Reg t) -> expr t
  | Eread_rf : forall n t, member $\Phi$ (Regfile n t) -> V (Int n) -> expr t
  (* operations on Booleans *)
  | Emux : forall t, V B -> V t -> V t -> expr t
  | Eandb : V B -> V B -> V B | ... 
  (* operations on words *)
  | Eadd : forall n, V (Int n) -> V (Int n) -> expr (Int n) | ... 
  (* operations on tuples *)
  | Efst : forall l t, V (Tuple (t::l)) -> expr t | ...
  
  Inductive scope (A : Type): Type :=
  | Send : A -> scope A
  | Sbind : forall (t: ty), expr t -> (V t -> scope A) -> scope A. 
  
  Inductive write : mem -> Type :=
  | WR : forall t, V t -> V B -> write (Reg t)
  | WRF : forall n t, V t -> V (Int n) -> V B ->  write (Regfile n t). 
       
  Definition effects := DList.T (option $\circ$ write) $\Phi$. 
  Definition block t := scope (V B * V t *  effects).         
End t.
Definition Block $\Phi$ t := forall V, block $\Phi$ V t.
\end{coq}
  \caption{RTL programs with three-adress code expressions}
  \label{fig:rtl}
\end{figure}

\subsubsection{Semantics.} We now turn to define the semantics of our RTL
language. 
First, we endow closed expressions with a denotation function (in the
same way as we did at the source level, except that it is not a
recursive definition).
Note that we instantiate the variable parameter of \coqe{expr} with
the function $\denotety{.}$, effectively tagging variables with their
denotations.

\begin{mcoq}
Variable $\Gamma$: $\denote{\Phi}$. 
Definition eval_expr (t : ty) (e : expr $\denotety{.}$ t) : $\denotety{.}$:=
  match e with
  | Evar t v => v
  | Einput t v => DList.get v $\Gamma$
  | Eread  t v =>  DList.get v $\Gamma$
  | Eread_rf n t v adr => Regfile.get (DList.get v $\Gamma$) adr
  | Emux t b x y => if b then x else y 
  | Eandb a b => andb a b | ...
  | Eadd n a b => Word.add a b  | ...
  | Efst l t e => Tuple.fst e | ...
  end. 
\end{mcoq}

\noindent The denotation of telescopes is a simple recursive function that
evaluates bindings in order and applies an arbitrary function on the
final (closed) object.
\begin{mcoq}
Fixpoint eval_scope {A B} (F : A -> B) (T : scope $\denotety{.}$ A) : B :=
match T with 
| Send X => F X
| Sbind t e cont => eval_scope F (cont (eval_expr t e))
end.   
\end{mcoq}
The final piece that we need is the denotation that corresponds to the
\coqe{write} type. This function takes as argument a single effect,
the initial state of this memory location  and either returns a new
state for this memory location, or returns \coqe{None}, meaning that
location is left in its previous state.
\begin{mcoq}
Definition eval_effect (m : mem) : option (write $\denotety{.}$ m) -> $\denotemem{m}$ -> option $\denotemem{m}$ := ... 
\end{mcoq}
Using all these pieces, it is quite easy to define what is the final
next-state function. 
\begin{mcoq}
Definition Next {t} {$\Phi$} ($\Gamma$: $\denote{\Phi}$) (B : Block $\Phi$ t) : option ($\denotety{t}$ * $\denote{\Phi}$) := ...
\end{mcoq}

\subsection{Compiling Fe-Si to RTL} 
Our syntactic translation from Fe-Si to RTL is driven by the fact that
our RTL language does not allow clashing assignments: syntactically,
each register and register file is updated by at most one \coqe{write}
expression.
With one wrinkle, we could say that we move to a language with
\emph{static single assignment}. 

\subsubsection{From control flow to data flow.}To do so, we have to
transform the control flow (the \coqe{Assert} and \coqe{OrElse}) of
Fe-Si programs into data-flow.
We can do that in hardware, because circuits are inherently parallel:
for instance, the circuit that computes the result of the conditional
expression \mbox{\coqe{e ? a : b}} is a circuit that computes the
value of \coqe{a} and the value of \coqe{b} in parallel and then uses
the value of \coqe{e} to select the proper value for the whole
expression.

\subsubsection{Administrative normal form.} Our first compilation pass
transforms Fe-Si programs into an intermediate language that
implements A-normal form. That is, we assign a name to every
intermediate computation.
In order to do so, we also have to resolve the control flow. To be
more specific, given an expression like
\begin{mcoq}
do x <- (A OrElse B); ... 
\end{mcoq}
we want to know statically to what value \coqe{x} needs to be bound
and when this value is \emph{valid}. 
In this particular case, we remark that if \coqe{A} yields a value
$v_A$ which is valid, then \coqe{x} needs to be bound to $v_A$; if
\coqe{A} yields a value that is invalid, then \coqe{x} needs to be
bound to the value returned by \coqe{B}. In any case, the value bound
in \coqe{x} is valid whenever the value returned by \coqe{A} or the
value returned by \coqe{B} is valid.

More generally, our compilation function takes as argument an
arbitrary function, and returns a telescope that binds three values:
(1) a \emph{guard}, which denotes the validity of the following
components of the tuple; %
(2) a \emph{value}, which is bound by the telescope to denote the value
that was returned by the action; %
(3) a list of \emph{nested effects}, which are a lax version of the
effects that exist at the \coqe{RTL} level.

The rationale behind these nested effects is to represent trees of
conditional blocks, with writes to state-holding elements at the
leaves. (Using this data type, several paths in such a tree may lead
to a write to a given memory location; in this case, we use a notion
of program order to discriminate between clashing assignments.)

\subsubsection{Linearizing the effects} Our second compilation pass
flattens the nested effects that were introduced in the first pass.
The idea of this translation is to associate two values to each
register: a \emph{data} value (the value that ought to be written) and
a \emph{write-enable} value. The data value is committed (i.e., stored)
to the register if the write-enable Boolean is true.
Similarly, we associate three values to each register-file: a data value, an
address, and a write-enable. The data is stored to the field of the
register file selected by the address if the write-enable is true.  

The heart of this translation is a \coqe{merge} function that takes
two writes of the same type, and returns a telescope that
encapsulates a single \coqe{write}: 
\begin{mcoq}
Definition merge s (a b : write s): scope (option (write s)) := ...   
\end{mcoq}
%
For instance, in the register case, given $(v_a,e_a)$ (resp. $(v_b,
e_b)$) the data value and the write-enable that correspond to
\coqe{a}, the write-enable that corresponds to the merge of \coqe{a}
and \coqe{b} is $e_a || e_b$, and the associated data value is
\mbox{$e_a~?~v_a : v_b$}.

\subsubsection{Moving to RTL.} The third pass of our compiler translates
the previous intermediate language to RTL, which amounts to a simple
transformation into three-address code. This transformation simply
introduces new variables for all the intermediate expressions that
appear in the computations. 

\subsection{Lightweight optimizations}
We will now describe two optimizations that we perform on programs
expressed in the RTL language. 
The first one is a syntactic version of common sub-expression
elimination, intended to reduce the number of bindings and introduce
more sharing. 
The second is a semantic common sub-expression elimination that aims
to reduce the size of the Boolean formulas that were generated in the
previous translation passes. 

\subsubsection{Syntactic common-subexpression elimination.}
We implement CSE with a simple recursive traversal of RTL
programs. (Here we follow the overall approach used by
Chlipala~\cite{DBLP:conf/popl/Chlipala10}.)

Contrary to our previous transformations that were just ``pushing
variables around'' for each possible choice of variable representation
\coqe{V}, here we need to tag variables with their symbolic values,
which approximate the actual values held by variables.
Then, CSE goes as follows. We fold through a telescope and maintain a
mapping from symbolic values to variables. For each binder of the
telescope, we compute the symbolic representation of the expression
that is bound. 
If this symbolic value is already in the map, we avoid the creation of
an extraneous binder. Otherwise, we do create a new binder, and extend
our association list accordingly.

\subsubsection{Using BDDs to reduce Boolean expressions.}
Our compilation process introduces a lot of extra Boolean
variables. We use BDDs to implement semantic common-subexpression
elimination. We implemented a BDD library in Coq; and we use it to
annotate each Boolean expression of a program with an approximation of
its runtime value, i.e, a pointer to a node in a BDD.
Our use of BDDs boils down to hash-consing: it enforces that Boolean
expressions that are deemed equivalent are shared.

The purpose of this pass is to simplify the extraneous boolean
operations that were introduced by our compilation passes. In order to
simplify only the Boolean computations that we introduced, we could
use two different kinds of Booleans (the ones that were present at the
source level and the others); and use our simplification pass only on
the latter.

\subsection{Putting it all together}
In the end, we prove that our whole compiler that goes from Fe-Si to
RTL and implements the two lightweight optimizations described above
is correct. That is, we prove that that next-step functions at the
source level and the target level are compatible.

\begin{twolistings}
  \begin{coq}
Variable ($\Phi$: list mem) (t : ty). 
Definition fesic  (A : Fesi.Action $\Phi$ t) : RTL.Block $\Phi$ t :=
  let x := IR.Compile $\Phi$  t a in
  let x := RTL.Compile $\Phi$ t x in 
  let x := CSE.Compile $\Phi$ t x in  
  BDD.Compile $\Phi$ t x.
\end{coq}
&
\begin{coq}
Theorem fesic_correct A :
forall ($\Gamma$ : $\denote{\Phi}$), 
Front.Next $\Gamma$ A = 
RTL.Next $\Gamma$ (fesic A).

$ $
\end{coq}
\end{twolistings}

\section{Design and verification of a sorter core}
We now turn to the description of a first hardware circuit implemented
and proved correct in Fe-Si. 
A \emph{sorting network}~\cite{DBLP:books/mg/CormenLRS01} is a
parallel sorting algorithm that sorts a sequence of values using only
compare-and-swap operations, in a data-independent way. This makes it
suitable for a hardware implementation.

Bitonic sorters for sequences of length $2^n$ can be generated using
short and simple algorithmic descriptions. Yet, formally proving their
correctness is a challenge that was only partially solved in two
different lines of previous work.
First, sorter core generators were studied from a hardware design
perspective in Lava~\cite{DBLP:conf/charme/ClaessenSS01}, but formal
proof is limited to circuits with a fixed size -- bounded by the
performances of the automated verification tools. 
Second, machine-checked formal proofs of bitonic sort were performed
e.g., in Agda~\cite{DBLP:conf/types/BoveC04}, but without a
connection with an actual hardware implementation. 
Our main contribution here is to implement such generators and to
propose a formal proof of their correctness.

More precisely, we implemented a version of bitonic sort as a regular
Coq program and proved that it sorts its inputs. This proof follows
closely the one described by Bove and
Coquand~\cite{DBLP:conf/types/BoveC04} -- in Agda -- and amounts to
roughly 1000 lines of Coq, including a proof of the so-called
\mbox{0-1~principle}\footnote{That is, a (parametric) sorting network
  is valid if it sorts all sequences of 0s and 1s.}.

Then, we implemented a version of bitonic sort as a Fe-Si program,
which mimicked the structure of the previous one. We present
side-by-side the Coq implementation of \coqe{reverse} in
Fig.~\ref{fig:reverse}.
The version on the left-hand side can be seen as a specification: it
takes as argument a sequence of $2^n$ inputs (represented as a
complete binary tree of depth $n$) and reverses the order of this
sequence.
The code on the right-hand side implements part of the
connection-pattern of the sorter. More precisely, it takes as input a
sequence of input variables and builds a circuit that outputs this
sequence in reverse order. 

Next, we turn to the function that is at the heart of the bitonic
sorting network.
A bitonic sequence is a sequence $(x_i)_{0 \le i < n}$ whose
monotonicity changes fewer than two times, i.e.,
$$ x_0 \le \cdots \le x_k \ge \cdots x_n, \text{with } 0 \le k < n $$
or a circular shift of such a sequence.
Given a bitonic input sequence of length $2^n$, the left-hand side
\coqe{min_max_swap} returns two bitonic sequences of length $2^{n-1}$,
such that all elements in the first sequence are smaller or equal to
the elements in the second sequence. 
The right-hand side version of this function builds the corresponding
comparator network: it takes as arguments a sequence of input
variables and returns a circuit. 

We go on with the same ideas to finish the Fe-Si implementation of
bitonic sort. The rest of the code is unsurprising, except that it
requires to implement a dedicated bind operation of type
\begin{mcoq}
forall (U: ty) n,  Var (domain n) -> (T n -> action [] Var U) -> action [] Var U. 
\end{mcoq}
that makes it possible to recover the tree structure out of the
result type of a circuit (\coqe{domain n}).

\begin{figure}[t]
  \centering
\begin{twolistings}
\begin{coq}
(* Lists of length $2^n$ represented as trees *)
Inductive tree (A: Type): nat -> Type :=
| L : forall x : A, tree A 0
| N : forall n (l r : tree A n), tree A (S n). 
$ $
Definition leaf  {A n} (t: tree A 0) : A := ...
Definition left  {A n} (t: tree A (S n)) : tree A n := ...
Definition right {A n} (t: tree A (S n)) : tree A n := ...

Fixpoint reverse {A} n (t : tree A n) :=
match t with 
| L x => L x
| N n l r => 
  let r := (reverse n r) in 
  let l := (reverse n l) in 
  N n r l
end.

$ $

Variable cmp: A -> A -> A * A.

Fixpoint min_max_swap {A} n : 
  forall (l r : tree A n), tree A n * tree A n :=
match n with 
| 0 => fun l r => 
  let (x,y) := cmp (leaf l) (leaf r) in (L x, L y)
| S p => fun l r => 
  let (a,b) := min_max_swap p (left l) (left r) in 
  let (c,d) := min_max_swap p (right l) (right r) in 
  (N p a c, N p b d)
end. 

...

Fixpoint sort n : tree A n -> tree A n := ...
\end{coq}
& $\quad$
\begin{coq}
Variable A : ty.        
Fixpoint domain n := match n with 
| 0 => A
| S n => (domain n) $\otimes$ (domain n)
end. 

Notation T n := (tree (expr Var A) n). 
Notation C n := action nil Var (domain n). 

Fixpoint reverse n (t : T n) : C n  :=
match t with 
| L x => ret x
| N n l r => 
  do r <- reverse n r;
  do l <- reverse n l;
  ret [tuple r, l]
end.

Notation mk_N x y := ([tuple x,y]).

Variable cmp : Var A -> Var A 
$\qquad\qquad\qquad$ -> action nil Var (A $\otimes$ A).
Fixpoint min_max_swap n : 
  forall (l r : T n), C (S n) :=
match n  with 
| 0 => fun l r => 
  cmp (leaf l) (leaf r)
| S p => fun l r => 
  do a,b <- min_max_swap p (left l) (left r);
  do c,d <- min_max_swap p (right l) (right r); 
  ret ([tuple mk_N a c, mk_N b d])
end.

...

Fixpoint sort n : T n -> C n := ...
\end{coq}
\end{twolistings}
  
  \caption{Comparing the specification and the Fe-Si implementation}
  \label{fig:reverse}
\end{figure}
We are now ready to state (and prove) the correctness of our sorter
core. We chose to settle in a context where the type of data that we
sort are integers of a given size, but we could generalize this proof
to other data types, e.g., to sort tuples in a lexicographic order.
We side-step the presentation of some of our Coq, to present the final
theorem in a stylized fashion.
\begin{theorem}
Let $I$ be a sequence of length $2^n$ of integers of size $m$. The
circuit always produces an output sequence that is a sorted permutation of $I$.
\end{theorem}
(Note that this theorem states the correctness of the Fe-Si
implementation against a specification of sorted sequences that is
independent of the implementation of the sorting algorithm in the
left-hand side of Fig.~\ref{fig:reverse}; the latter only serves as a
convenient intermediate step in the proof.)


\subsubsection{Testing the design}
Finally, we indulge ourselves and test a design that was formally
proven, using a stock Verilog simulator~\cite{iverilog}. We set the
word size and the number of inputs of the sorter, and we generate the
corresponding Verilog code. Unsurprisingly, the sorter core sorts its
input sequence in every one of our test runs.

\section{Verifying a stack machine}
The circuit that was described in the previous section is a simple
combinational sorter: we could have gone one step further in this
verification effort and pipelined our design by registering the output
of each compare-and-swap operator. However, we chose here to describe
a more interesting design: a hardware implementation of a simple stack
machine, inspired by the IMP virtual
machine~\cite{Leroy-Marktoberdorf-09}, i.e., a small subset of the
Java virtual machine.

Again, we proceed in two steps: first, we define a specification of
the behavior of our stack machine; second, we build a Fe-Si
implementation and prove that it behaves as prescribed. 
The instruction set of our machine is given in Fig.~\ref{fig:stack},
where we let $x$ range over identifiers (represented as natural
numbers) and $n,\delta$ range over values (natural numbers).
The state of the machine is composed of the code (a list of
instruction), a program counter (an integer), a variable stack (a list
of values), and a store (a mapping from variables to values). The
semantics of the machine is given by a one-step transition relation in Fig.~\ref{fig:stack}.
Note that this specification uses natural numbers and lists in a
pervasive manner: this cannot be faithfully encoded using finite-size
machine words and register files. 
For simplicity reasons, we resolve this tension by adding some dynamic
checks (that do not appear explicitly on Fig.~\ref{fig:stack}) to the
transition relation to rule out such ill-defined behaviors.

\begin{figure}
  \centering
  \begin{twolistings}
    \begin{tabular}{rcll}
i & ::=   & \texttt{const $n$                     }\\
  & $|$   & \texttt{var   $x$                              }\\
  & $|$   & \texttt{setvar  $x$                            }\\
  & $|$   & \texttt{add                                        }\\
  & $|$   & \texttt{sub                                        }\\
  & $|$   & \texttt{bfwd $\delta$             }\\
  & $|$   & \texttt{bbwd $\delta$              }\\
  & $|$   & \texttt{bcond $c$ $\delta$ }\\
 \\
  & $|$   & \texttt{halt                                       }\\
      \end{tabular}
& \qquad 
      \begin{tabular}{ll}
$C \vdash pc,\sigma,s \to pc+1, n :: \sigma,s$ & \text{if $C(pc)$ = \texttt{const $n$}} \\
$C \vdash pc,\sigma,s \to pc+1, s(x) :: \sigma,s$ & \text{if $C(pc)$ = \texttt{var $x$}} \\
$C \vdash pc,v::\sigma,s \to pc+1, \sigma,s[x \leftarrow v]$ & \text{if $C(pc)$ = \texttt{setvar $x$}} \\
$C \vdash pc,n_2::n_1::\sigma,s \to pc+1, (n_1+n_2)::\sigma,s$ & \text{if $C(pc)$ = \texttt{add}} \\
$C \vdash pc,n_2::n_1::\sigma,s \to pc+1, (n_1-n_2)::\sigma,s$ & \text{if $C(pc)$ = \texttt{sub}} \\
$C \vdash pc,\sigma,s \to pc+1+\delta, \sigma,s$ & \text{if $C(pc)$ = \texttt{bfwd $\delta$}} \\
$C \vdash pc,\sigma,s \to pc+1-\delta, \sigma,s$ & \text{if $C(pc)$ = \texttt{bbwd $\delta$}} \\
$C \vdash pc,n_2::n_1::\sigma,s \to pc+1+\delta, \sigma,s$ & \text{if $C(pc)$ = \texttt{bcond c $\delta$} and $c~n_1~n_2$} \\
$C \vdash pc,n_2::n_1::\sigma,s \to pc+1, \sigma,s$ & \text{if $C(pc)$
  = \texttt{bcond c $\delta$} and $\neg (c~n_1~n_2)$} \\
\texttt{no reduction}
      \end{tabular}
\end{twolistings}
  \caption{Instruction set and transition relation of our stack machine}
  \label{fig:stack}
\end{figure}

The actual Fe-Si implementation is straightforward. The definition of
the internal state is depicted below: the stack is implemented using a
register file, and a stack pointer; the store is a simple register
file; the code is implemented as another register file that is
addressed by the program counter.

\begin{twolistings}
\begin{coq}
Variable n : nat. 
Notation OPCODE := (Tint 4).  
Notation VAL := (Tint n). 

Definition INSTR := OPCODE $\otimes$ VAL.  
$ $
\end{coq}
&
\begin{coq}
Definition $\Phi$ : state := [
Tregfile n INSTR;       (* code *)
Treg VAL;               (* program counter *)
Tregfile n VAL;         (* stack *)
Treg VAL;               (* stack pointer *)
Tregfile n VAL          (* store *)]. 
\end{coq}
\end{twolistings}

The actual implementation of the machine is unsurprising: we access
the code memory at the address given by the program counter; we
case-match over the value of the opcode and update the various
elements of the machine state accordingly. 
For the sake of space, we only present the code for the \texttt{setvar
  $x$} instruction below.

\begin{twolistings}
\begin{coq}
Definition pop :=
do sp <- ! SP;       
do x <- read STACK [: sp - 1];
do _ <- SP ::= sp - 1;
ret x.    
\end{coq}
&
\begin{coq}
Definition Isetvar pc i := 
do v <- pop; 
do _ <- write REGS [: snd i <- v];
PC ::= pc + 1.
$ $
\end{coq}
\end{twolistings}

\subsubsection{Correctness.} We are now ready to prove that our
hardware design is a sound implementation of its specification. 
First, we define a logical relation that relates the two
encodings of machine state (in the specification and in the
implementation), written $\equiv$. 
Then, we prove a simulation property between related states.
\begin{theorem}
  Let $s_1$ be a machine state as implemented in the specification and
  $m_1$ the machine state as implemented in the circuit, such that
  $s_1 \equiv m_1$.
  If $s_1$ makes a transition to $s_2$, then $m_1$ makes a transition
  to $m_2$ such that $s_2 \equiv m_2$.
\end{theorem}
Note that we do not prove completeness here: it is indeed the case
that our implementation exhibits behaviors that cannot be mapped to
behaviors of the specification. 
In this example, the specification should be regarded as an
abstraction of the actual behaviors of the implementation, which could
be used to reason about the soundness of programs, either written by
hand or produced by a certified compiler.

\subsubsection{Testing the design}
Again, we compiled our Fe-Si design to an actual Verilog
implementation. We load binary blobs that correspond to test programs
in the code memory, and run it while monitoring the content of given
memory locations. This gives rise to an highly stylized way of
computing e.g., the Fibonacci sequence.

\section{Comparison with related work}\label{sec:rw}
Fe-Si marries hardware design, functional programming and inductive
theorem proving.
We refer the reader to Sheeran~\cite{DBLP:journals/jucs/Sheeran05} for
a review of the use of functional programming languages in hardware
design, and only discuss the most closely related work.

\medskip

Lava~\cite{Bjesse98lava:hardware} embeds parametric circuit generators
in Haskell. 
Omitting the underlying implementation language, Lava can be
faithfully described as a subset of Fe-Si, with two key differences.
First, Lava features a number of layout primitives, which makes it
possible to describe more precisely what should be the hardware
layout, yielding more efficient FPGA implementation. We argue that
these operators are irrelevant from the point of view of verification,
and could be added to Fe-Si if needed.
Second, while Lava features ``non-standard'' interpretations of
circuits that make it possible to prove the correctness of fixed-size
tangible representation of circuits, our embedding of Fe-Si in Coq
goes further: it makes it possible to prove the correctness of
parametric circuit generators.

\medskip

Bluespec~\cite{bluespec} is an industrial strength hardware
description language based on non-deterministic guarded atomic
actions. A program is a set of rewrite rules in a Term Rewriting
System that are non-deterministically executed one at a time. 
To implement a Bluespec program in hardware, the Bluespec compiler
needs to generate a deterministic schedule where one or more rules
happen each clock-cycle.
Non-determinism makes it possible to use Bluespec both as an
implementation language and as a specification language.
Fe-Si can be described as a deterministic subset of Bluespec. 
We argue that deterministic semantics are easier to reason with, and
that we can use the full power of Coq as a specification language to
palliate our lack of non-determinism in the Fe-Si language.
Moreover, more complex scheduling can be implemented using a few
program combinators~\cite{DBLP:conf/memocode/DaveAP07}: we look
forward to implementing these in Fe-Si. 

\medskip 

Richards and Lester~\cite{DBLP:journals/isse/RichardsL11} developed a
shallow-embedding of a subset of Bluespec in PVS. While they do not
address circuit generators nor the generation of RTL code, they proved
the correctness of a three-input fair arbiter and a two-process
implementation of Peterson's algorithm that complements our case
studies (we have not attempted to translate these examples into Fe-Si).

Slind et al~\cite{DBLP:journals/fac/SlindOIG07} built a compiler that
creates correct-by-construction hardware implementations of arithmetic
and cryptographic functions that are implemented in a synthesisable
subset of HOL. Parametric circuits are not considered. 

Centaur Technology and the University of Texas have developed a formal
verification framework~\cite{DBLP:conf/memocode/SlobodovaDSH11} that
makes it possible to verify RTL and transistor level code. 
They implement industrial level tools tied together in the ACL2
theorem prover and focus on hardware validation (starting from
existing code). By contrast, we focus on high-level hardware
synthesis, with an emphasis on the verification of parametric designs.

\section{Conclusion}
Our compiler is available on-line along with our examples as
supplementary material~\cite{fesi}.
The technical contributions of this paper are
\begin{itemize}
\item a novel use of PHOAS as a way to embed domain-specific languages
  in Coq;
\item a toy compiler from a simple hardware description language to
  RTL code;
\item machine checked proofs of correctness for some simple hardware
  designs. 
\end{itemize}

This work is intended to be a proof of concept: much remains to be
done to scale our examples to more realistic designs and to make our
compiler more powerful (e.g., improving on our current
optimizations). Yet, we argue that it provides an economical path to
certification of parameterized hardware designs.

\bibliography{synthesis}
\end{document}